\documentclass[11pt]{article}
\usepackage{moriond2000,epsfig}

\def\Real{{\rm I\mathchoice{\kern-0.70mm}{\kern-0.70mm}{\kern-0.65mm}%
  {\kern-0.50mm}R}}

\font \bolditalics = cmmib10
\def\bx#1{\leavevmode\thinspace\hbox{\vrule\vtop{\vbox{\hrule\kern1pt
        \hbox{\vphantom{\tt/}\thinspace{\bf#1}\thinspace}}
      \kern1pt\hrule}\vrule}\thinspace}

\def \vc #1{{\textfont1=\bolditalics \hbox{$\bf#1$}}}

\def\thetag{{\vc \theta}}

\def\be{\begin{equation}}
\def\ee{\end{equation}}
\def\bea{\begin{eqnarray}}
\def\eea{\end{eqnarray}}

\begin{document}
\vspace*{4cm}
\title{COSMIC SHEAR WITH THE CFHT}

\author{L. VAN WAERBEKE$^{1,2}$, Y. MELLIER$^{2,3}$, T. ERBEN$^4$, J.C.
CUILLANDRE$^5$, F. BERNARDEAU$^6$,
R. MAOLI$^{2,3}$, E. BERTIN$^{2,3}$, H.J. Mc CRACKEN$^7$, O. LE FEVRE$^7$, B.
FORT$^2$,
M. DANTEL-FORT$^3$, B. JAIN$^{8}$, P. SCHNEIDER$^4$}
 
\address{$^1$ Canadian Institut for Theoretical Astrophysics, 60 St 
Georges
Str., Toronto, M5S 3H8 Ontario, Canada.\\
   $^2$ Institut d'Astrophysique de Paris. 98 bis, boulevard
Arago. 75014 Paris, France. \\
   $^3$ Observatoire de Paris. DEMIRM. 61, avenue de
l'Observatoire.  75014 Paris, France.\\
   $^4$ Max Planck Institut fur Astrophysiks, Karl-Schwarzschild-Str. 1,
Postfach 1523,
D-85740 Garching, Germany. \\
   $^5$ Canada-France-Hawaii-Telescope, PO Box 1597, Kamuela, Hawaii 96743,
USA\\
   $^6$ Service de Physique Th\'eorique. C.E. de Saclay. 91191 Gif sur 
Yvette
Cedex, France.\\
   $^7$ Laboratoire d'Astronomie Spatiale, 13376 Marseille Cedex 12, 
France\\
   $^8$ Dept. of Physics, Johns Hopkins University, Baltimore, MD 21218, 
USA\\
}

\maketitle\abstracts{
We present preliminary results of our cosmic shear survey currently in progress
at the Canada-France-Hawaii Telescope (CFHT). We analysed 1.7 sq. degrees of high
quality data (seeing below 1 arcsec), out of which we were able to measure a
significant correlation of galaxy shape-orientation over several arcmin scale.  
We present measurements of the variance of the shear $\langle \gamma^2\rangle$
and of the correlation functions $\langle e_r(\theta)e_r(0)\rangle$,
$\langle e_t(\theta)e_t(0)\rangle$, and we show that the signal is consistent with
gravitational lensing by large scale structures predictions. The level of
residual systematics after Point Spread Function correction is discussed and
shown to be small compared to our measured signal. We outline several possible
future evolutions of our work by using additional data of our lensing survey now extended
to 9 sq. degrees.
}

\section{Introduction}

Since the pioneering works by \cite{b91,k92,m91,v96},
gravitational lensing by large scale
structure has been recognized as a potential powerful probe of the mass
distribution in the Universe at scales from 1 arcmin to several degrees. The
effect we are looking for consists of a small stretching of distant galaxies
by the foreground matter, with an amplitude of 1-5 percents of distortion. Since the
induced distortion is very small compared to the intrinsic ellipticities of the
galaxies ($\sim 0.3$), it can be observed only statistically by measuring the averaged
orientation and elongation of several galaxies. Unfortunately telescope imaging
defects (like optical distortion, tracking errors, etc...) induce also a
significant coherent elongation of the galaxies, very similar to
the gravitational lensing effect we wish to measure. Therefore the most
challenging part of the analysis is the correction of the intrinsic telescope defects
(which makes the Point Spread Function anisotropic) down to a level smaller than the
expected lensing signal. This problem prevented the detection of the
lensing effect in the earlier searches \cite{m94,v84,v95}. Recently a significant
measurement of the galaxy alignment was found thanks to the improvement of the
image quality \cite{s98}, but the analyzed fields were small,
and yet it was not clear what sort of mass distribution was responsible for the
lensing effect.

The fundamental importance of a clean PSF correction pushed many groups to
design specific methods of image analysis \cite{b95,k00,k95,k99,r00,v97}. During
the preparation of
our cosmic shear program, we tested intensively the Kaiser, Squires and
Broadhurst (KSB) correction scheme on
highly realistic simulations \cite{e00}. We found that the KSB method can easily
reach the
$1\%$ accuracy PSF correction even for severely deteriorated PSF's as we
obtain in real observations.
A $1\%$ accuracy might be not sufficient to measure the gravitational lensing
effect at the degree scale, but at the arcmin scale the gravitational lensing
signal is enhanced by the non-linear evolution of the large-scale-structures
\cite{j97}, and the variance of the shear $\langle \kappa^2\rangle$ can be as
high as $3-5\%$. We are therefore confident that the present day technology is
enough to detect the cosmic shear effect, even if significant improvements remain to
be done to fully exploit the scientific case.

Four independent groups almost simultaneously reported a significant detection
\cite{b00,kwl00,v00,wit} and a new detection is coming with  VLT data
\cite{m00,mhere00}. Here,
we present our own detection \cite{v00}, and discuss different aspects of the PSF
correction and of the statistical measurement. Next we compare the measurements with the
predictions and discuss their consistency. We show how our cosmic shear
program will hopefully improve the situation very shortly: we outline the new
measurements and the new tests of PSF correction we plan to do next.

\begin{table*}
\caption{List of the fields. Most of the exposures were taken in the I band
at CFHT. The total area is $1.7$ deg$^2$, and the 8 fields are uncorrelated.
}
\label{fields}
\begin{center}
\begin{tabular}{|c|c|c|c|c|c|c|c|}
\hline
Target & Name & Camera & Used area & Filter & Exp. time & Period & seeing\\
\hline
F14P1 & F1 & CFH12K & 764 ${\rm arcmin}^2$ & V& 5400 sec. & May 1999 & 
0.9"\\
F14P2 & F2 & CFH12K & 764 ${\rm arcmin}^2$ & V & 5400 sec. & May 1999 & 
0.9"\\
F14P3 & F3 & CFH12K & 764 ${\rm arcmin}^2$ & V & 5400 sec. & May 1999 & 
0.9"\\
CFDF-03 & F4 & UH8K & 669 ${\rm arcmin}^2$ & I & 17000 sec. & Dec. 1996 &
0.75"\\
SA57 & F5 & UH8K & 669 ${\rm arcmin}^2$ & I &12000 sec. & May 1998 & 0.75"\\
A1942 & F6 & UH8K & 573 ${\rm arcmin}^2$ & I & 10800 sec. & May 1998 & 
0.75"\\
F02P1 & F7 & CFH12K & 1050 ${\rm arcmin}^2$ & I & 9360 sec. & Nov. 1999 & 
0.8"\\
F02P4 & F8 & CFH12K & 1050 ${\rm arcmin}^2$ & I & 7200 sec. & Nov. 1999 & 
0.9"\\
\hline
\end{tabular}
\end{center}
\end{table*}

\section{From the data set to the galaxy catalogues}

Table \ref{fields} show the list of the fields used in our cosmic shear detection.
Despite the relatively large differences in the filter and the exposure time, the mean
redshift depth is approximately $1$ according to the deepest
spectroscopic surveys done so far \cite{coh99}, with a dispersion probably very large ($\sim 1$). It is impossible to
give a more accurate determination of redshifts as we do not have enough colors for this.
A complete scientific interpretation of our signal would require the missing
redshift information, but in the early stages of the work we were more interested in a
{\it detection} of the cosmic shear effect rather than its scientific exploitation. As
discussed in the last Section, the scientific analysis requires some crucial issues to be
addressed first.

The total field covers about $6300~{\rm arcmin}^2$ and has a number density of galaxies of
about $n_g\simeq 30~{\rm gal/arcmin}^2$. However, after a proper weighting of the galaxies
their effective number density is about half (details are in the original paper \cite{v00}).
Our procedure of shape measurement using IMCAT \footnote{See Nick Kaiser's
home page at {\sf http://www.ifa.hawaii.edu/$\sim$kaiser/}.}
is described in details
elsewhere \cite{e00,mhere00,v00} so it is unnecessary to give it here, therefore we assume
that we have already a catalogue of PSF corrected ellipticities for the galaxies.

\begin{figure}
\begin{center}
\psfig{figure=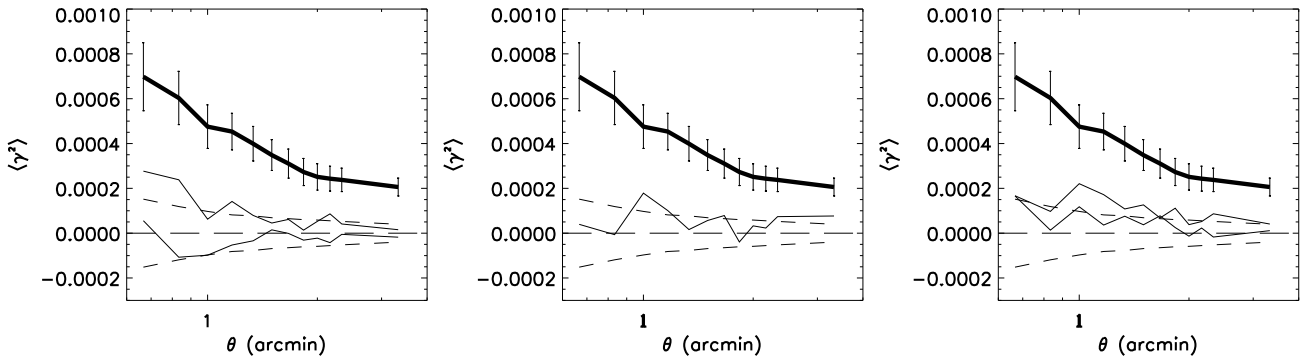,height=1.9in}
\caption{The thick solid line show the measured $\langle \gamma^2\rangle$, and the error bars
are calculated using randomized ellipticity orientations. The dashed lines show the
$\pm 1 \sigma$ levels. From left to right: (a) the thin solid lines are
$\langle \gamma^2\rangle$ measured in bins of $N$ galaxies, where the galaxies are chosen
with respect to the star ellipticity strength $e_1$ and $e_2$. The number of galaxies is then
converted into a fictive {\it scale} which is used to make the plot. (b) The $N$ galaxies
are chosen with respect to the optical distortion amplitude. (c) The $N$ galaxies are chosen
with respect to the CCD lines and columns.
\label{syst_correct_HR}}
\end{center}
\end{figure}

\section{From the galaxy catalogues to the cosmological parameters}

\subsection{What is the relevant information?}

The easiest quantities to measure for the detection of the cosmic shear effect
are the variance of the
shear $\langle \gamma^2\rangle$ and the ellipticity correlation functions. For a simplified
cosmological model (zero cosmological constant, power law power spectrum with slope $n$
and normalization $\sigma_8$, and a single redshift plane $z_s$), we can show that
\cite{j97}:

\begin{equation}
\langle \gamma^2\rangle^{1/2}\simeq 0.01\sigma_8\Omega_0^{0.75}z_s^{0.75}\left(
{\theta\over {\rm arcmin}}\right)^{\left({n+2\over 2}\right)},
\label{variance}
\end{equation}
where $\Omega_0$ is the density of the Universe and $\theta$ the measurement scale in
arcmin. There are similar relations for the correlation functions, but we can define
several types of correlation functions. Here we are interested in $\langle e_t(\theta)e_t(0)
\rangle$, $\langle e_r(\theta)e_r(0)\rangle$ and $\langle e_t(\theta)e_r(0)\rangle$, where
$e_t$ and $e_r$ are the tangential and the radial component of the shear respectively:

\begin{eqnarray}
e_t&=&-\gamma_1 \cos (2\theta_{gal})-\gamma_2 \sin(2\theta_{gal})\nonumber\\
e_r&=&-\gamma_2 \cos (2\theta_{gal})+\gamma_1 \sin(2\theta_{gal}),
\label{corrfunc}
\end{eqnarray}
where $\gamma_i$ are the Cartesian components of the shear and $\theta_{gal}$ is the position
angle of the pair of galaxies. The correlation functions mentioned above have a particular
interest because of the specific signatures induced from weak lensing by large scale structures
\cite{m91}: from the scalar nature of the gravity we can show that
$\langle e_t(\theta)e_r(0)\rangle$ should vanish, that $\langle e_t(\theta)e_t(0)\rangle$
is positive and $\langle e_r(\theta)e_r(0)\rangle$ should become negative for a finite range
of scale.

\subsection{Measurements and amplitude of the residual systematics}

\begin{figure}
\begin{center}
\psfig{figure=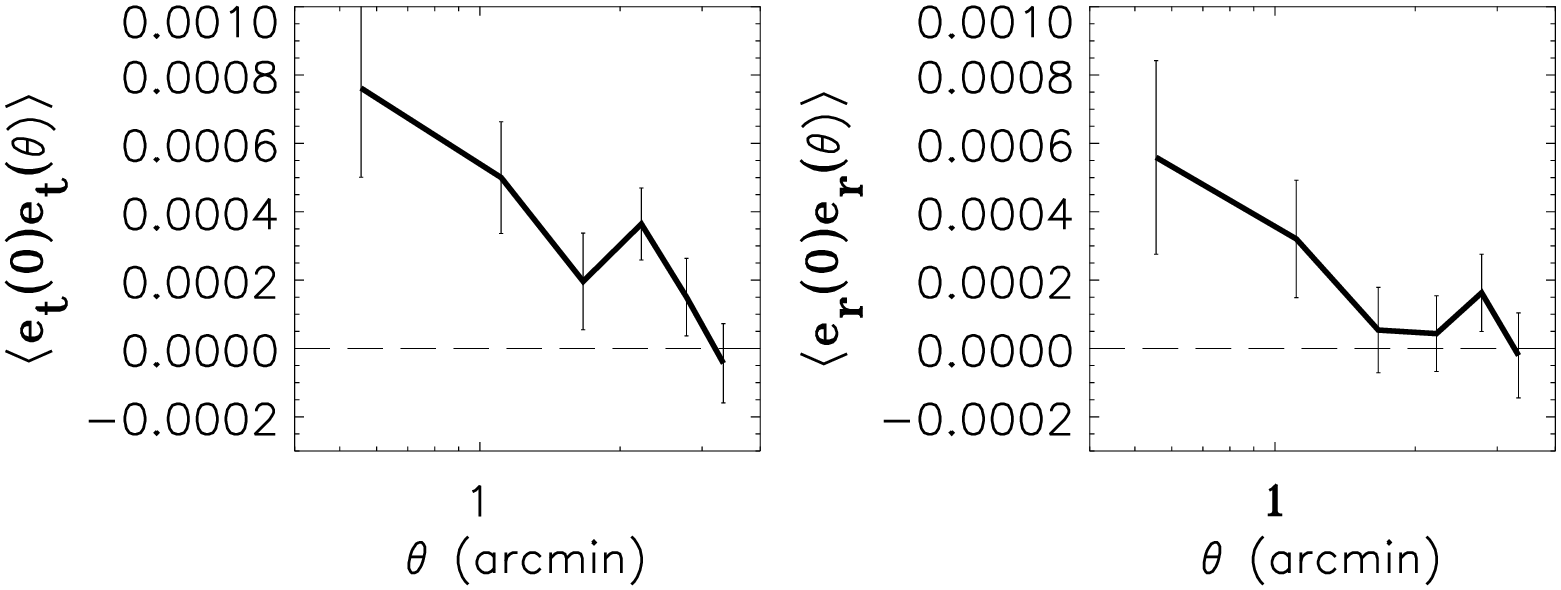,height=1.9in}
\caption{Measured ellipticity correlation functions $\langle e_t(\theta)e_t(0)\rangle$ (left)
and $\langle e_r(\theta)e_r(0)\rangle$ (right) .
\label{correl_fct_HR}}
\end{center}
\end{figure}

We assumed that we already have a catalogue of galaxy ellipticities $\bf e$ corrected from
the PSF anisotropy. We call $\sigma_\epsilon$ the ellipticity dispersion.
From a set of $N$ galaxies at positions $\thetag_k$ with ellipticities
$e_\alpha(\thetag_k)$ we can built an estimate of the variance of the shear at the
position $\thetag_i$

\begin{equation}
E[\gamma^2(\thetag_i)]={\displaystyle \sum_{\alpha=1,2}} \left({1\over
N}{\displaystyle \sum_{k=1}^N} e_\alpha
(\theta_k)\right)^2,
\label{varestimator}
\end{equation}
whose the ensemble average is $\langle E[\gamma^2(\thetag_i)] \rangle=\sigma_\epsilon^2/N
+\langle\gamma^2\rangle$. The term $\sigma_\epsilon^2/N$ can be removed either by measuring
it on randomized catalogues, or by choosing and unbiased estimate equal to Eq.(
\ref{varestimator}) without the diagonal elements.

An estimate of the correlation function $E[e_i(0)e_j(r)]$ (where $i$ and $j$ are either
$t$ and/or $r$) is built by summing $e_i(0)e_j(r)$ for all the possible pairs of galaxies
separated by a distance $r$

Figure \ref{syst_correct_HR} shows the measured variance of the shear in our survey. The
signal (thick line) exhibits the characteristic power law scale dependence as predicted 
from Eq.(\ref{variance}), and the amplitude of the signal has the correct order of magnitude.
However, as we shall see in the next Section, the cosmic variance and the error bars are
too large to put tight constraints on the cosmology.

We have shown in our original paper \cite{v00} that the corrected ellipticity of the
galaxies are uncorrelated with the ellipticity of the stars, which demonstrates the low
level of residual systematics present in the catalogues \footnote{However we found
a constant bias of $\langle e_1\rangle\simeq -0.01$ for all the galaxies which we corrected
for. The source of the bias is still unclear.}. We have reproduced in Figure
\ref{syst_correct_HR} (thin solid lines) the variance of the shear measured when
the galaxies are picked up
according to the local star ellipticity instead of taking the galaxies falling into a
given smoothing window. Each estimate of the shear variance is done using
Eq.(\ref{varestimator}),
where the $N$ galaxies have in common a similar PSF anisotropy amplitude on the
data. It is then easy to convert $N$ to a {\it scale} $\theta$, since the number density
of galaxies is roughly constant. The resulting shear variance is shown as the thin solid
lines on Figure \ref{syst_correct_HR} for three cases of possible source of residual
systematics (see caption for details). It shows that any residual systematic cannot be
due to star ellipticity, optical distortion and CCD frame alignment.

A robust test of the gravitational lensing origin of the signal is to measure the
correlation function as indicated above. Figure \ref{correl_fct_HR} shows the two
relevant correlation functions $\langle e_t(r)e_t(0)\rangle$ and
$\langle e_r(r)e_r(0)\rangle$. The third one, not shown here,
$\langle e_t(r)e_r(0)\rangle$ is zero as we expect from the scalar origin of
the gravitational field. This results from the fact that systematics are almost non-existent
in the galaxy catalogue, which is a strong supports for the cosmic origin of our signal. However 
the survey size is still too small to have a low noise measurement of these quantities, and
to extract the useful cosmological information. Instead, we shall see this detection as
a success showing the feasibility of the cosmic shear searches.

\subsection{Cosmological constraints}

\begin{figure}
\begin{center}
\psfig{figure=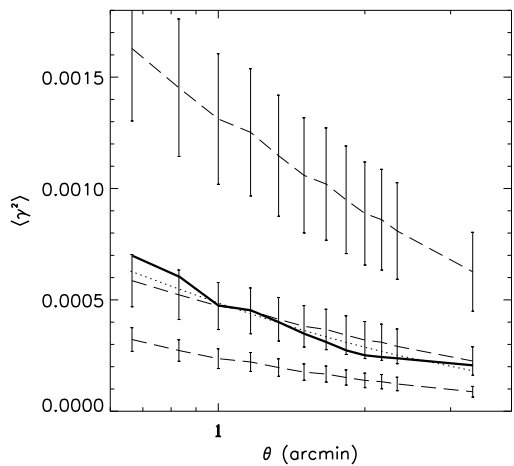,height=2.5in}
\caption{The thick solid line is drawn from Figure \ref{correl_fct_HR} without the measurement
error bars. The dashed lines show the prediction of
$\langle \gamma^2\rangle$ from ray-tracing simulations 
and the cosmic variance
obtained from several realizations. From top to bottom: $(\Omega_0,\sigma_8)=
(1.,1.); (0.3,0.85); (0.3, 0.6)$. The source redshift is 1 and the power spectrum CDM-type.
\label{cosmic_constraint}}
\end{center}
\end{figure}

Although it is hard to calculate the cosmic variance of the measured
quantities at small scale, we can use ray tracing numerical simulations in order
to compare our measurement with some realistic scenario. Such numerical simulations
are available \cite{j00} and provide several realizations for a set of cosmological models,
thus allowing a first analysis of the cosmic variance. Figure \ref{cosmic_constraint} is a
comparison of our shear variance measurement (thick solid line) with three different
models: $(\Omega_0,\sigma_8)=(1.,1.); (0.3,0.85); (0.3, 0.6)$ from top to bottom (see
the caption for details). The error bars show the cosmic variance measured out of several
realizations for each model. Although we can only marginally reject the two extreme
models, this figure shows that the data are already in favor of cluster normalized
models, as suggested by the simplified analytical estimate Eq.(\ref{variance}).

\section{Future prospects}


The aim of the above Section was to present a condensed view of our recent cosmic shear
detection with the first stream of data obtained at CFHT. All the relevant technical
details can be found elsewhere \cite{e00,v00}; here we emphasized the
control of the systematics
to a reasonable level and that the signal is consistent with the gravitational lensing
predictions, although we are not yet in good position to put tight constraints on the
cosmology. However, we now have 9 sq. degres in I (instead of 2), ready for scientific
analysis, of comparable image quality
of the data discussed here, and the numerical simulations \cite{j00,v99} indicate
that this is enough to make a very significant measurement of various lensing statistics.

It is worthwhile to ask what can be done next. The next step is the confirmation of the
cosmological nature of the signal over many different aspects. There are two
 areas with significant room for improvement:

1-\underline{Beating the systematics}: following the same strategy in our cosmic shear
paper, we can
search for systematics by binning the data in various ways. With 9 sq. degrees we can bin
the data in a much refined way than we did before. We can study the alignment of the
galaxies with the stars in bins of magnitude, size or other parameter. We can also split
the survey into smaller parts and analyze independently the different parts.

2-\underline{Search for specific gravitational lensing signatures}: this was done with
the correlation
function, but we can do much more with 9 sq. degrees. Of course the signal-to-noise of the
correlation function will be much better, and we can study its shape as a function of size
and magnitude of the galaxies. The variance of the shear measured with another filter than
top-hat will be possible: for instance the Map statistic \cite{sv98,v99} is known to have a
specific angular dependence if induced by gravitational lensing. With a 9 sq. degrees
survey we can also measure the skewness of the convergence \cite{j00,v99}, which
is a direct probe of the cosmological parameters \cite{b97,j97}. Such measure
would require accurate mass reconstruction in non-trivial topologies (non-straight edges
and holes in the field), and preliminary studies show that such reconstructions
give stable and unbiased mass maps. Peak statistics \cite{jv00} is also accessible
with a good signal-to-noise with such a survey size. Measurement of the shear
power spectrum will be done in order to check that the power is distributed over the
different scales like what weak lensing theory predicts (and do not dominate at one specific
scale for instance).

It has recently been suggested that galaxies might have an intrinsic ellipticity
correlation \cite{cm00,h00}. Although this could also be tested using specific
lensing statistics (yet to be developed), we can completely get rid of the effect
by measuring the
correlation between different source redshift planes. Since our full survey will be 16
sq. degrees in four colors, we hope to have the luxury to work with many different
redshift planes (using photometric redshifts) and to extract without ambiguity the
cosmic shear information {\it and} the intrinsic correlations.

\section*{Acknowledgments}

We are very gratefull to the organisers for this interesting and lively meeting.
We are particulary gratefull to Dmitri Pogosyan who was forced to attend all the
talks and succeeded brillantly in the hard task to make a non-conventional
and accurate summary talk. This work was supported by the TMR Network
"Gravitational Lensing: New Constraints on Cosmology and the Distribution of Dark
Matter" of the EC under contract No. ERBFMRX-CT97-0172. We thank the TERAPIX data
center for providing infinite computing resources for the reduction of the
CFH12K and UH8K data.


\section*{References}
\begin{thebibliography}{99}
\bibitem{b00} Bacon, D., Refregier, A., Ellis, R., 2000, astro-ph/0003008.
\bibitem{b97} Bernardeau, F., Van Waerbeke, L., Mellier, Y., 1997, A\&A, 322, 1.
\bibitem{b91} Blandford R., Saust, A., Brainerd, T., Villumsen, J., 1991, MNRAS, 251, 600.
\bibitem{b95} Bonnet, H., Mellier Y., 1995, A\&A, 303, 331.
\bibitem{coh99} Cohen, J. G., Hogg, D., W., Blandford, R., Cowie, L.,
Hu, E. , Songaila, A., Shopbell, P., Richberg, K., 1999,
astro-ph/9912048.
\bibitem{cm00} Croft, R., Metzler, C., 2000, astro-ph/0005384.
\bibitem{e00} Erben, T., Van Waerbeke, L., Bertin, E., et al., 2000, A\& A, astro-ph/0007021.
\bibitem{h00} Heavens, A., Refregier, A., Heymans, C., 2000, astro-ph/0005269.
\bibitem{j97} Jain, B., Seljak, U., 1997, ApJ, 484, 560.
\bibitem{j00} Jain, B., Seljak, U., White, S., 2000, ApJ, 530, 547.
\bibitem{jv00} Jain, B., Van Waerbeke, L., 2000, ApJL, 530, L1.
\bibitem{k92} Kaiser, N., 1992, ApJ, 388, 272.
\bibitem{k00} Kaiser, N., 2000, ApJ, 537, 555.
\bibitem{k95} Kaiser, N., Squires, G., Broadhurst, T., 1995, ApJ, 449, 460.
\bibitem{kwl00} Kaiser, N., Wilson, G., Luppino, 2000, astro-ph/0003338.
\bibitem{k99} Kuijken, K., 1999, A\&A, 352, 355.
\bibitem{m00} Maoli, R., et al., 2000, {\it in preparation}.
\bibitem{mhere00} Maoli, R., et al., {\rm these proceedings}.
\bibitem{m91} Miralda-Escude, J., 1991, ApJ, 380, 1.
\bibitem{m94} Mould, J., Blandford, R., Villumsen, J., et al., 1994, MNRAS, 271, 31.
\bibitem{r00} Rhodes, J., Refregier, A., Groth, E., 2000, ApJ, 536, 79.
\bibitem{sv98} Schneider, P., Van Waerbeke, L., Jain, B., Kruse, K., 1998, MNRAS, 296, 873.
\bibitem{s98} Schneider, P., Van Waerbeke, L., Mellier, Y., et al., 1998, A\& A, 333, 767.
\bibitem{v84} Valdes, F., Jarvis, J.F., Tyson, J.A., 1984, ApJ, 271, 431.
\bibitem{v99} Van Waerbeke, L., Bernardeau, F., Mellier, Y., 1999, A\&A, 342, 15.
\bibitem{v00} Van Waerbeke, L., Mellier, Y., Erben, T., et al., 2000, A\& A, 358, 30.
\bibitem{v97} Van Waerbeke, L., Mellier, Y., Schneider, et al., 1997, A\&A, 317, 303.
\bibitem{v95} Villumsen, J., 1995, astro-ph/9507007.
\bibitem{v96} Villumsen, J., 1996, MNRAS, 281, 369.
\bibitem{wit}Wittman, D., Tyson, J.A., Kirkman, D., et al., 2000, Nature, 405, 143.
\end{thebibliography}
\end{document}